\documentclass{PoS}

\usepackage{enumerate}
\usepackage{amsmath,amssymb}  
\usepackage{bm}               
\usepackage{amscd}            
\usepackage{graphicx}         
\usepackage{epsfig}
\usepackage{hhline,multirow}  
\usepackage{dcolumn}          
\usepackage{wrapfig}
\usepackage[dvipsnames]{xcolor}
\usepackage[normalem]{ulem}
\usepackage{mathtools}

\usepackage{slashed}
\usepackage{arydshln} 

\newcommand{\psl}{p\!\!\!\slash} 
\newcommand{\ksl}{k\!\!\!\slash}

\newcommand{\CF}{ \text{CF}}

\newcommand{\nbar}{{\overline n}}

\newcommand{\Tr}{{\rm Tr}}

\title{Testing collinear factorization in a spectator model with mass corrections}

\ShortTitle{Testing collinear factorization in a spectator model with mass corrections}

\author{\speaker{Juan V. Guerrero} and Alberto Accardi \\
	Hampton U. and Jefferson Lab, USA\\
	E-mail: \email{juanvg@jlab.org}}  

\abstract{In perturbative QCD, the masses of the hadrons involved in high energy reactions can usually be neglected. However, in the case of Kaon production in electron-proton collisions at low (and not so low) beam energies this may not be a good approximation. In particular, a recent proposal to include hadron masses in theoretical calculations shows how these Hadron Mass Corrections (HMCs) can explain a large discrepancy observed in measurements performed at the HERMES and COMPASS experiments. In this talk, we present preliminary results of a spectator model calculation designed to test the range of validity of the approximations needed in the proposed factorization scheme. We focus on inclusive DIS scattering as a first step towards the analysis of HMCs in semi-inclusive processes.}

\FullConference{XXVII International Workshop on Deep-Inelastic Scattering and Related Subjects (DIS2019)\\
		8-12 April 2019\\
		Torino, Italy}

\begin{document}

\section{Introduction}

Collinear Parton Distribution Functions (PDFs) in QCD encode information regarding the longitudinal quark and gluon structure of hadrons and can be accessed through hard scattering reactions.
This takes advantage of QCD factorization theorems, such as Collinear Factorization (CF), which are usually formulated in the asymptotically large limit of some physical scale, {\it e.g.} the photon virtuality $Q^2$ in the case of Deep Inelastic Scattering (DIS).
However, experiments at low beam energy, such as at Jefferson Lab, involve low photon virtualities, and require theoretical control of $\mu^2 / Q^2$ kinematical power corrections in addition to dynamical higher twist effects. For example, in inclusive DIS $\mu=M$ is the target mass, and Semi Inclusive DIS receives corrections also from the mass of the observed hadron.

These ``Hadron Mass Corrections" (HMCs) have been recently explored in Ref.~\cite{Accardi:2009md,Guerrero:2015wha}, and can even affect relatively high-energy experiments such as HERMES and COMPASS, possibly explaining the apparent large discrepancy between their measurements of integrated kaon multiplicities \cite{Guerrero:2015wha,Guerrero:2017yvf}. At Jefferson Lab, mass effects become large also for pion production. 
In this talk we present a model calculations designed to test the validity ot the sub-asymptotic kinematic approximations needed in the Hadron Mass Correction Scheme~\cite{Guerrero:2017yvf}. We initially focus on inclusive DIS in order to avoid complications due to the non trivial interplay of initial and final state kinematics in semi-inclusive processes discussed in Ref.~\cite{Guerrero:2015wha}.
\section{DIS in a spectator model}

\begin{figure}[tb]
	\centering
	\includegraphics[width=9cm]{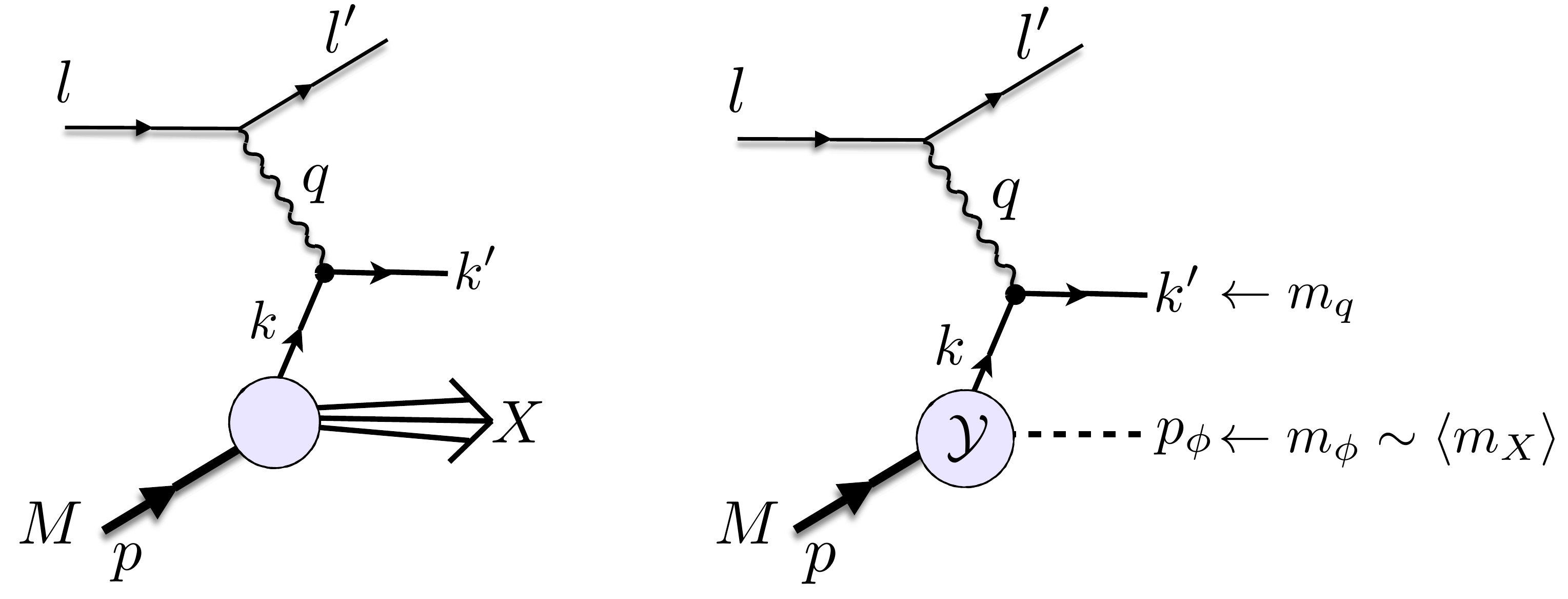}	
	\caption{Electron-proton DIS kinematics
          {\it Left diagram}: real one, where $X$ is remmant of the target fragmentation. \textit{Right diagram}: spectator model, with remnant replaced by a scalar diquark.} 
	\label{Fig:DIS}
\end{figure}

The DIS kinematics is defined in Fig.~\ref{Fig:DIS} left. In the final state, the remnant $X$ and the recoil quark momentum $k'$ are not measured. However, we assume the identity of the latter can be experimentally determined, in analogy with measurement of the charm-tagged $F_2^c$ structure function, and we assume the quark mass $m_q$ to be known.

The four-momenta of the external particles and target quark can be parametrized in terms of light cone unit vectors $n$ and $\nbar$, with 
$n^2 = \nbar^2 = 0$ and $n\cdot\nbar=1$.
In the so called ``$(p,q)$ frame'' \cite{Accardi:2009md}, in which the 
target and virtual photon are coplanar, with zero tranverse momentum ($\boldsymbol{p_T}=\boldsymbol{q_T}=\boldsymbol{0}$)
\begin{align*}
p^\mu = p^+\, \nbar^\mu  
+ \frac{M^2}{2 p^+}\, n^\mu,	  \hspace*{0.5cm}	
q^\mu = - \xi p^+\, \nbar^\mu 
+ \frac{Q^2}{2\xi p^+}\, n^\mu,	  \hspace*{0.5cm}	
k^\mu = xp^+\, \bar{n}^\mu
+ \frac{k^2 + \bm{k}_\perp^2}{2 x p^+}\, n^\mu
+ k^\mu_\perp,				
\end{align*}
where
$
\xi \equiv -\frac{q^+}{p^+} =
\frac{2 x_B}{1 + \sqrt{1 + 4 x_B^2 M^2/Q^2}}
$
is the so-called Nachtmann scaling variable, $x_B=\frac{Q^2}{2 p\cdot q}$ is the Bjorken scaling variable, and $x = \frac{k^+}{p^+}$ is the light-cone momentum fraction carried by the parton~\footnote{The ``plus'' and ``minus'' components
$a^\mu$ are defined by
$a^+ = a \cdot n     = (a^0 + a^3)/\sqrt{2}$ and
$a^- = a \cdot \nbar = (a^0 - a^3)/\sqrt{2}$.}.

Our goal is to mimic and simulate the real electron-proton DIS  ocurring in nature with a simpler, calculable toy model. 
For this, we use an idealized field-theory \cite{Bacchetta:2008af,Moffat:2017sha} with a spin 1/2 particle representing a nucleon of mass $M$, an active quark of mass $m_q$, and a scalar diquark ``spectator''. At LO, the latter describes target fragmentation, which is a complex process in QCD, with a single particle of mass $m_\phi \sim \langle m_X \rangle$.
In this ``spectator model'', the nucleon, quark and the spectator interact through a vertex $\mathcal{Y}= i g(k^2) \mathbb{I}$,
with $g(k^2)$ a dipolar form factor:
\begin{equation}
g(k^2) = g \frac{k^2-m_q^2}{|k^2-\Lambda^2|^2}
\end{equation}
where $g$ is an appropriate coupling constant (playing no role in the present discussion) and the parameter $\Lambda$ cuts off ultraviolet divergences in $k^2 \gg \Lambda^2$  \cite{Bacchetta:2008af}. This cutoff also imposes a minimal length scale of $\mathcal{O}(1/\Lambda)$, and effectively simulates confinement in the nucleon target.

\begin{figure}
	\centering
	\includegraphics[width=12cm]{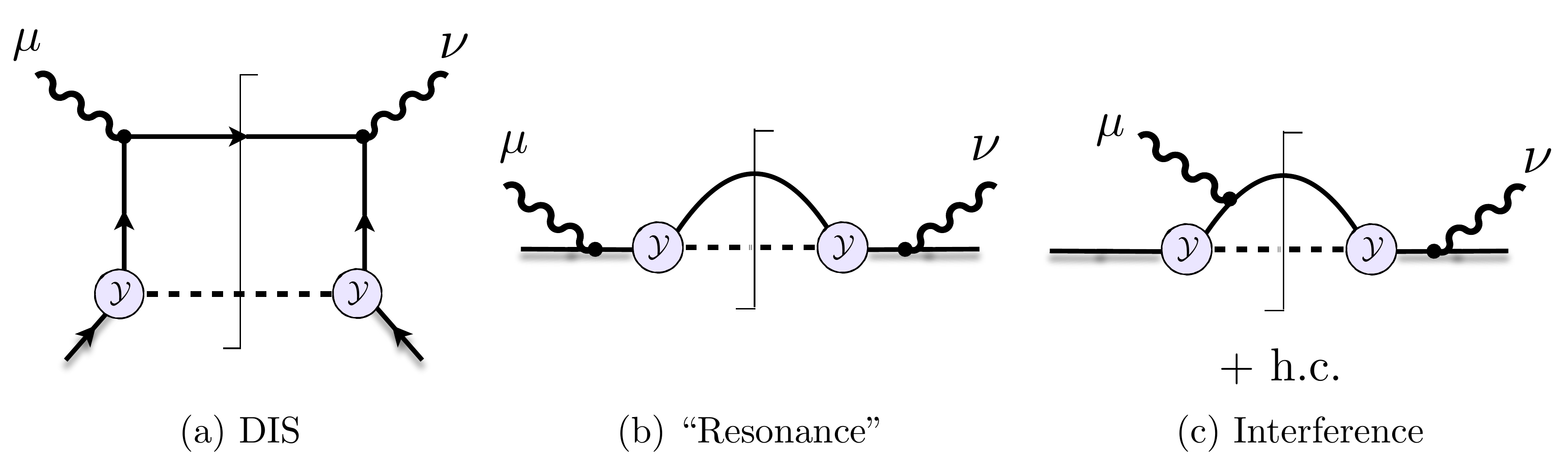}	
	\caption{Diagrams contributing to $ep$-scattering up to order $g^2$.} 
	\label{Fig:ep_scattering}
\end{figure}

In contrast to the real electron-proton scattering, in the spectator model it is possible to calculate the hadronic tensor in an exact way.  At the lowest order in $g$, this receives contributions from the 3 processes depicted in Fig.~\ref{Fig:ep_scattering}:
photon-quark scattering, which mimics DIS; photo-excitation of the proton with subsequent decay into a quark and a spectator; and interference between these two.

The contribution of each diagram in Fig.~\ref{Fig:ep_scattering} is not gauge invariant by itself, but the sum is. Since each diagram contains different physics, we can use parity invariant projectors~\cite{Aivazis:1993kh,Accardi:2008ne,Guerrero:2019xxx} to extract their individual gauge invariant structure functions; for example,
\begin{eqnarray}
F_{1, \text{g.i.}}^\text{DIS} &=&   \frac{1}{2} \Big(-\hat{g}^{\mu \nu} +\frac{\hat{p}^\mu \hat{p}^\nu}{\hat{p}^2}\Big)W_{\mu\nu}^{\text{DIS}}
\end{eqnarray}	
where $\hat{p}^\mu = p^\mu - \frac{p \cdot q}{q^2} q^\mu$, $\hat{g}^{\mu \nu} = g^{\mu\nu} - \frac{q^\mu q^\nu}{q^2} $,
and $W_{\mu\nu}^{\text{DIS}}$ is calculated from the diagram in Fig.~\ref{Fig:ep_scattering}a.
Here we focus on the DIS piece because collinear factorization is intended to provide a controlled approximation only to this contribution. The role of the other two diagrams is discussed in the talk's slides and in Ref.~\cite{Guerrero:2019xxx}.

\section{Collinear factorization for DIS in the spectator model}

In the model, it is also possible to calculate the structure function with the collinear approximation used to analyze real life DIS:
$
  F_1 \approx F_1^{\, \CF} = \int \frac{dx}{x} \, \mathcal{H}_1(x) q(x),
$
where $\mathcal{H}_1 = 2 \pi \, e_q^2\,  x \,  \delta(x-\bar{x})$ 
is the tree level hard scattering coefficient, and $q(x) = \includegraphics[height=1cm]{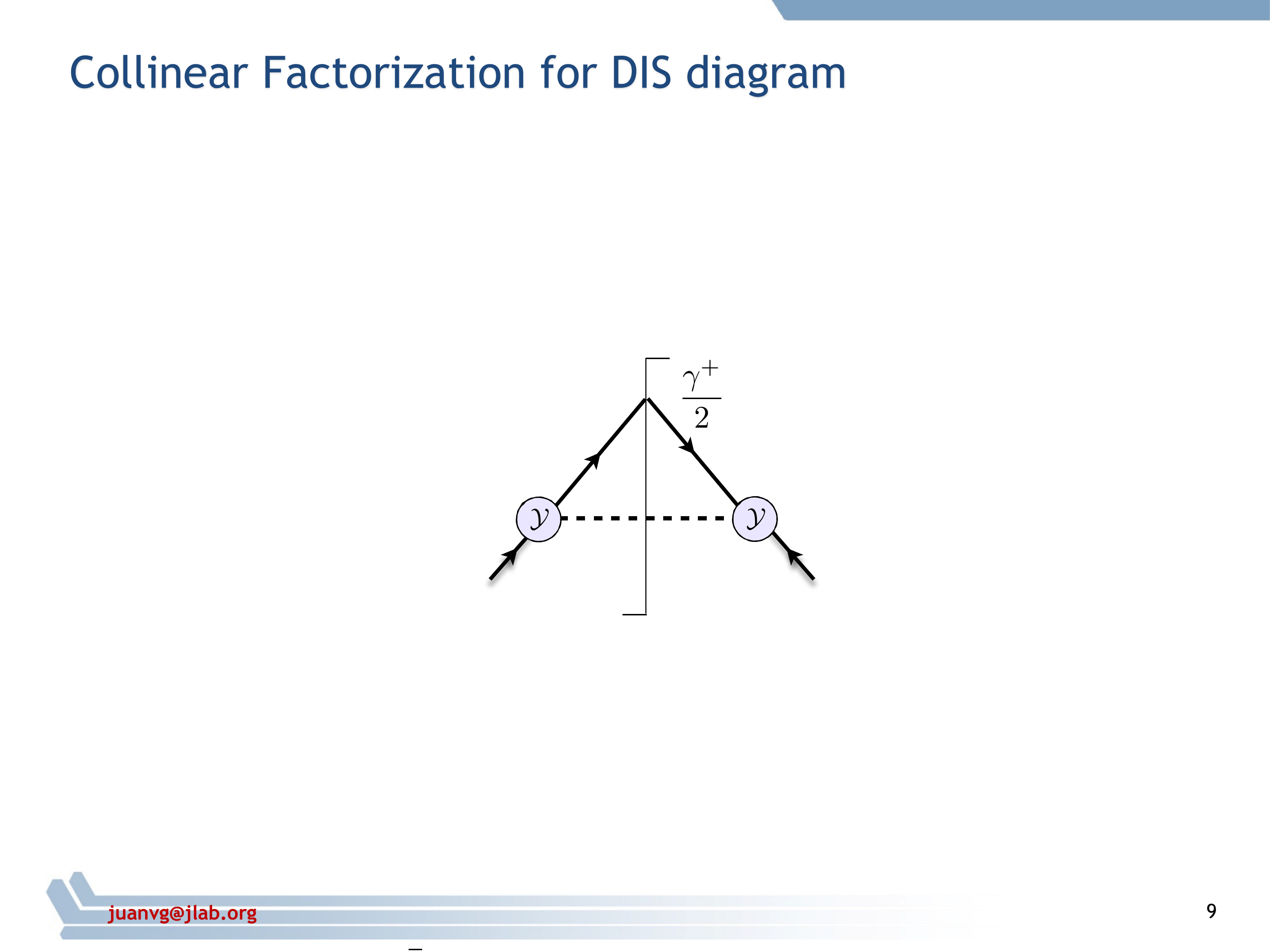}= \int dk^-d^2k_T   \frac{\Tr[(\psl+M)(\ksl+m_q)\frac{\gamma^+}{2}(\ksl+m_q)]}{(k^2-m_q^2)^2}$
is the PDF, also calculable in the model! 
The delta function is a consequence of four-momentum conservation in the hard scattering vertex and the top cut in Fig.~\ref{Fig:ep_scattering}(a). Finally,
\begin{equation}
F_{1}^{\text{DIS}}(x_B,Q^2,\mu) \approx F^{\, \CF}_1(x_B,Q^2,\mu) =  2\pi \, e_q^2\,\, q(\bar x,Q^2,\mu) \, ,
\label{eq:F1approx}
\end{equation}
where $\bar x$ approximates the light cone fraction for the incoming quark determined by the hard scattering kinematics in the full diagram. This reads
\begin{eqnarray}
  x
   &=& \xi \Bigg(1 + \frac{m_q^2+ k_T^2}{Q^2} - \frac{(m_q^2+ k_T^2)(k^2+ k_T^2)}{Q^4} + \mathcal{O}\Big(\frac{\mu^6}{Q^6}\Big) \Bigg) \ ,
\label{eq:x_sol}
\end{eqnarray}
and depends only on two mass scales, namely, the scattered transverse quark mass $m_{qT}^2 = m_q^2 + k_T^2$ and the ``light cone virtuality'' of the incoming parton, $v^2 = k^2 + k_T^2$, that measures how different from zero $k^-=v^2/(2k^+)$ is.

\begin{wrapfigure}[8]{R}{0.46\textwidth}
  \vskip-0.8cm
    \hspace*{-0.35cm}
    \begin{tabular}{ c | c  c | c }
      $\bar{x}$ & $M$ & $m_q$  & $k_T^2$  \\
      \hline
      $x_B$ & $0$ & $0$ & $0$ \\[5pt]
      $\xi$ & $\checkmark$ & $0$ & $0$ \\[5pt]
      $\xi\Big(1+\frac{m_q^2}{Q^2}\Big) \equiv \xi_q$ & $\checkmark$ & $\checkmark$ & $0$  \\[5pt]
      \hdashline 
      $\xi\Big(1+\frac{m_q^2 + \langle k_T^2 \rangle}{Q^2}\Big) \equiv \xi_q^{(T)}$ & $\checkmark$ & $\checkmark$ & $\checkmark\, \langle k_T^2 \rangle$ \\[5pt]
    \end{tabular}
    \caption{Kinematical approximations for $x$.}
    \label{eq:xapprox}
\end{wrapfigure}
In the full diagram the unobserved $k^2$ would be fixed by the botton cut in Fig.~\ref{Fig:ep_scattering}(a), but neither this nor the equally unobserved $k_T^2$ can be controlled in an inclusive process and we have to resort to a kinematic approximation to determine $x$. Nonetheless, we need only to worry about the latter, or rather only about $m_{qT}$, since $v^2$ (and therefore $k^2$) only contributes at $\mathcal{O}(1/Q^4)$ and is negligible in DIS kinematics. 
We can identify three (plus one) $x \approx \bar x$ approximations depending on the kinematic variables we decide to neglect: see the table in Fig.~\eqref{eq:xapprox},
where the approximation above the dashed line include only external, experimentally observabe quantities, and below the line we considered the most relevant internal variable, namely, $k_T^2$, albeit only on average.

Despite the fact that $v^2 = k^2 + k_T^2$ and $k_T^2$ are internal variables and cannot be directly measured in inclusive processes, we can compute their average values in the model:
$
\langle \mathcal{O} \rangle (x_B,Q^2) = \frac{\int^{k_{T,\text{max}}^2}_0  dk_T^2 dk^2 dx \, \mathcal{O}(x,,k_T^2,k^2) \mathcal{F}^\text{DIS}_1(x,k_T^2,k^2)}{\int^{k_{T,\text{max}}^2}_0 dk_T^2 dk^2 dx \, \mathcal{F}^\text{DIS}_1(x,k_T^2,k^2)}
\label{eq:avg_obs}
$, where $k_{T,max}^2$ is determined by the available invariant mass and the final state particle masses (the virtuality $k^2$ is determined at LO by the lower cut in Fig.~\ref{Fig:DIS}a), ${\cal O}(x,k_T^2,k^2)$ is a generic observable, and $\mathcal{F}^\text{DIS}_1$ is the unintegrated structure function. The dependence on $x_B$ and $Q^2$ are due to the external kinematics, which is left understood on the right hand side. In collinear factorization, momentum conservation in the transverse (and minus) directions is neglected in order to reduce the loop integration only to the  plus light cone direction, and the averages read
$
\langle \mathcal{O} \rangle_{\CF} (x_B,Q^2) = \frac{\int^{\infty}_0 dk_T^2 dk^2 dx \,  \mathcal{O}(\bar x,k_T^2,k^2) \mathcal{F}_1^{\, \CF}(x,k_T^2,k^2)}{\int^{\infty}_0 dk_T^2 dk^2 dx \,\mathcal{F}_1^{\, \CF}(x,k_T^2,k^2)}
\label{eq:avg_obs_col}
$ with no limit on $k_T^2$.

\begin{figure}
	\centering
	\includegraphics[height=4.75cm]{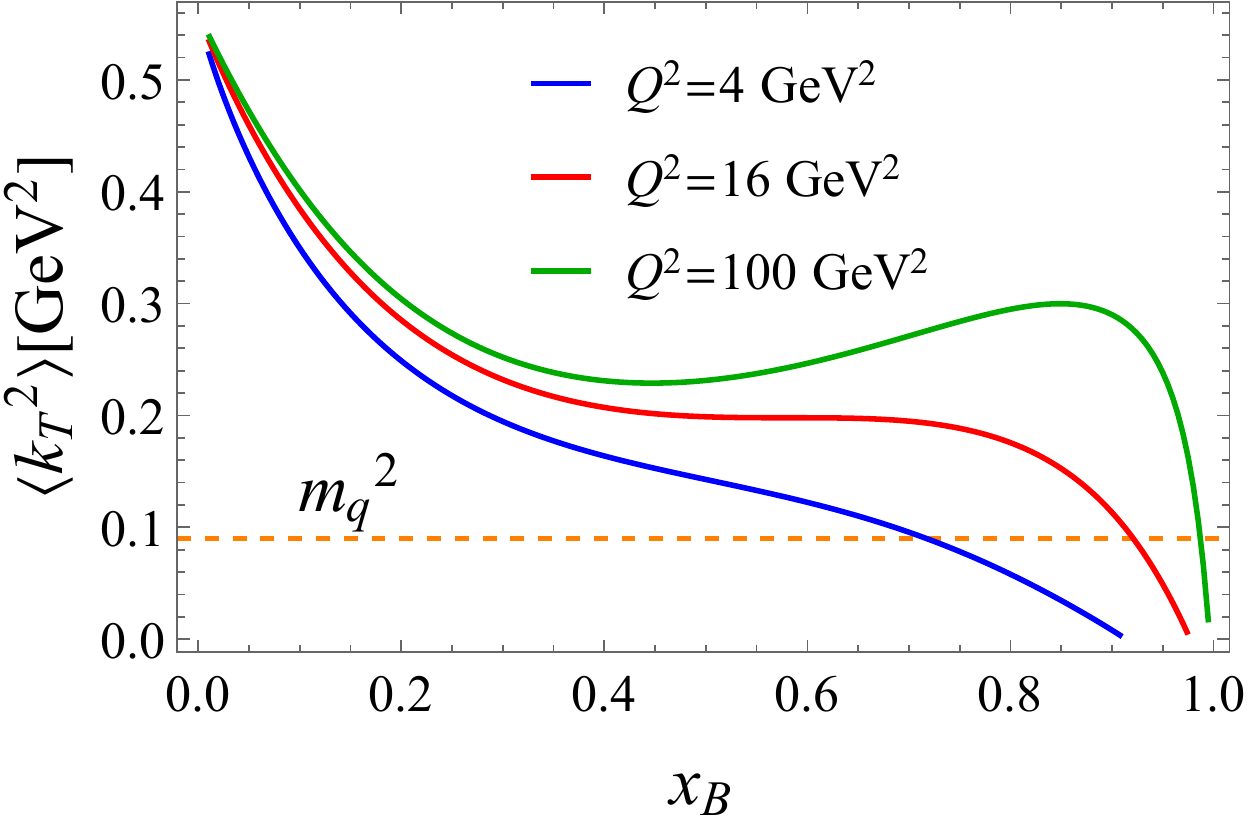}	
	\includegraphics[height=4.75cm]{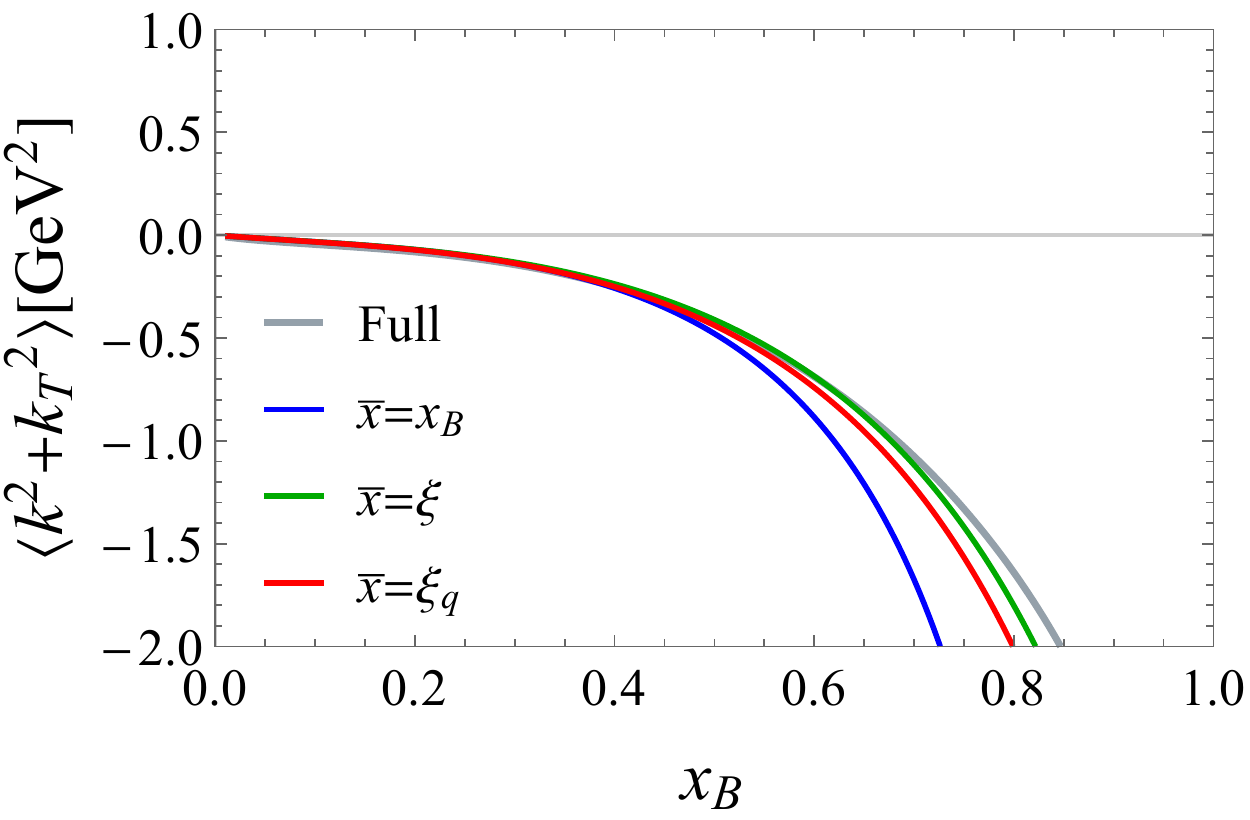}	
	\vskip-0.1cm
	\caption{Average ``unobserved'' kinematics of the incoming quark as a function of $x_B$. \textit{Left:} average $k_T^2$ in the full model (the orange dashed line indicates the value of $m_q^2$ for reference). \textit{Right:} average light cone virtuality $v^2=k^2+k_T^2$ in the full model, compared to various collinear kinematic approximations.} 
	\label{Fig:kt2_v2}
\end{figure}

In the left panel of Fig.~\ref{Fig:kt2_v2}, we show $\langle k_T^2 \rangle$ as a function of $x_B$ for several values of $Q^2$, and notice that $\langle k_T^2 \rangle \sim \mathcal{O}(m_q^2)$ is not a priori negligible in Eq.~\eqref{eq:x_sol}. In the right panel, we show the average light-cone virtuality $\langle k^2 + k_T^2 \rangle$, finding that at small values of $x_B$ the parton behaves like a collinear massless parton. It is also clear that the incoming quark virtuality is negative, $\langle k^2 \rangle < - \langle k_T^2 \rangle$, as it should for a bound quark. Thus, the typical parton model approximation $k^2 \sim 0$ can be quite inaccurate, and should rather be substituted with $k^- \sim 0$.

\section{Testing factorization}

We can now test the validity of the generalized collinear approximation \eqref{eq:F1approx} by comparing the exact calculation with the factorized $F_1^{\, \CF}$ for each kinematic approximation listed in Fig.~\ref{eq:xapprox} (we use $m_q = 0.3 {\rm\ GeV}$, $\Lambda = 0.609 {\rm\ GeV}$ and $m_\phi = 0.822 {\rm\ GeV}$, fitted to known PDFs in Ref.~\cite{Bacchetta:2008af}). 

\begin{figure}[b]
	\centering
	\includegraphics[height=4.75cm]{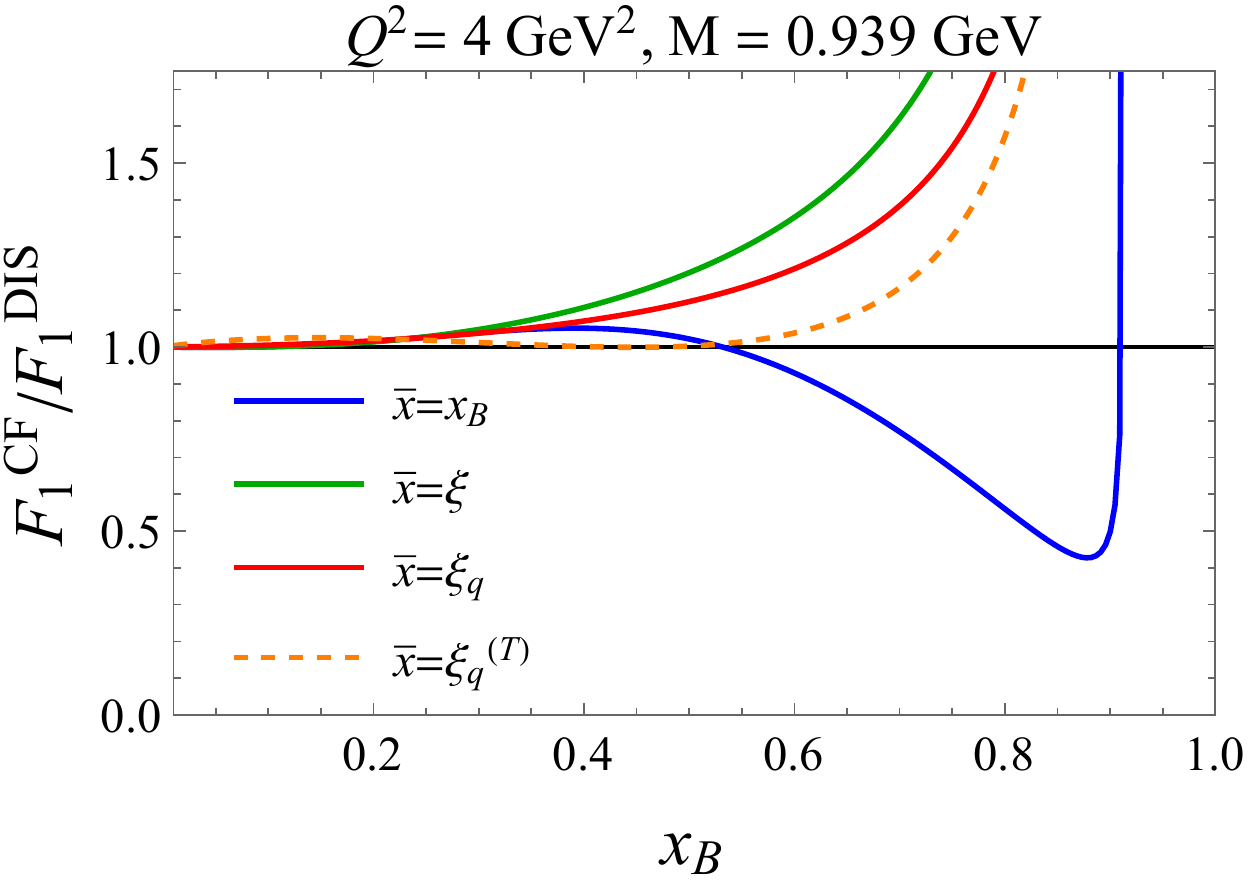}	
	\includegraphics[height=4.75cm]{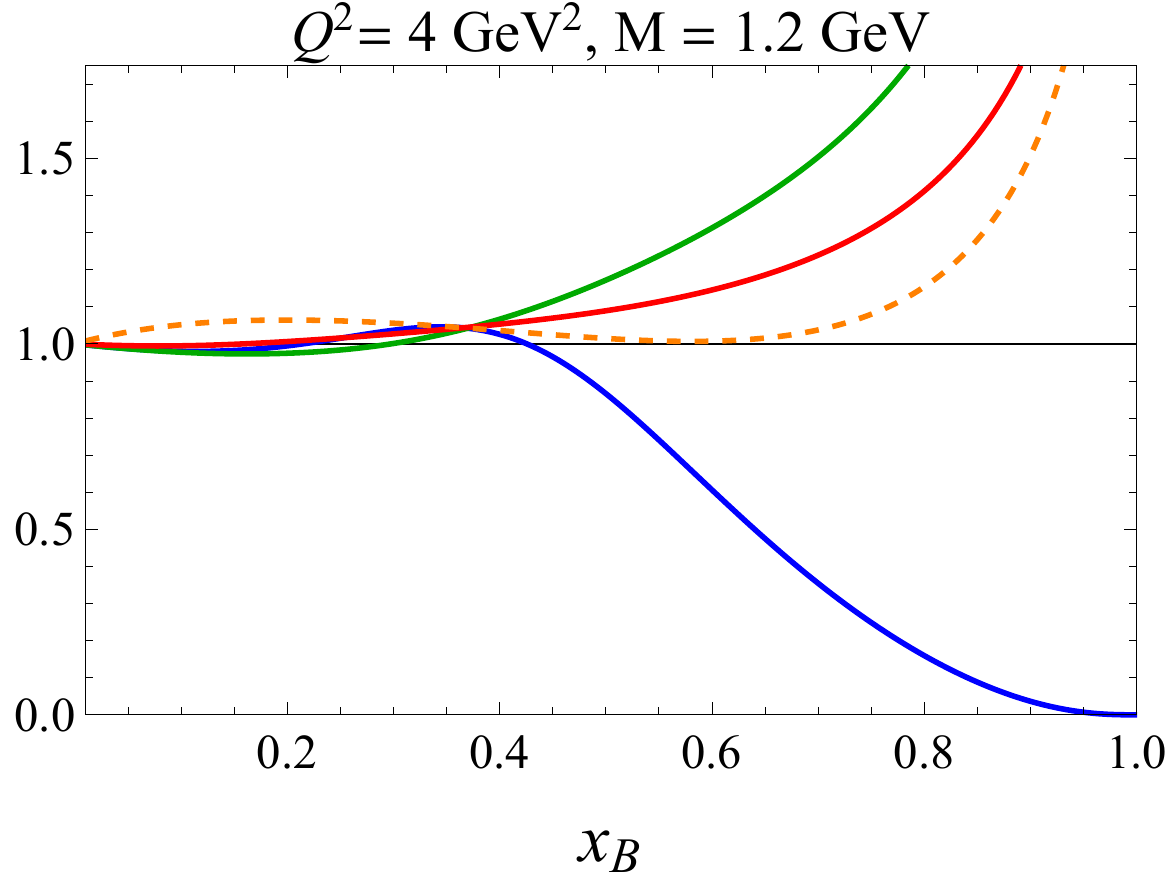}
	\vskip-0.1cm
	\caption{Ratio of collinear to full DIS structure functions.}
	\label{Fig:ratio_F1}
\end{figure}

In Fig.~\ref{Fig:ratio_F1}, we show the ratio of collinear to full $F_1$ structure functions for each proposed approximation in Table~\eqref{eq:F1approx}, and two different target mass values. The usual, asymptotic choice $x=x_B$ in general is not a good approximation, since it depends strongly on the target mass (something already observed in Ref.~\cite{Moffat:2017sha} for $k_T$-dependent structure functions). On the other hand, the choice $x=\xi_q$ provides the closest approximation to the full $F_1$ using only external variables.
The effect of the ``missing'' $k_T$ is shown by the dashed-orange line in Fig.~\ref{Fig:ratio_F1}: the additional $\langle k_T^2 \rangle/Q^2$ correction largely recovers the exact calculation. These transverse-motion-induced power corrections are however outside of the reach of our leading twist calculation, but can likely be handled  extending this to higher-twist~\cite{Qiu:1988dn}. Even so, we notice that factorization would break for $x_B \gtrsim 0.6$ because it does not respect momentum conservation in the transverse direction.

\begin{figure}[t]
  \centering
  \includegraphics[height=4.4cm]{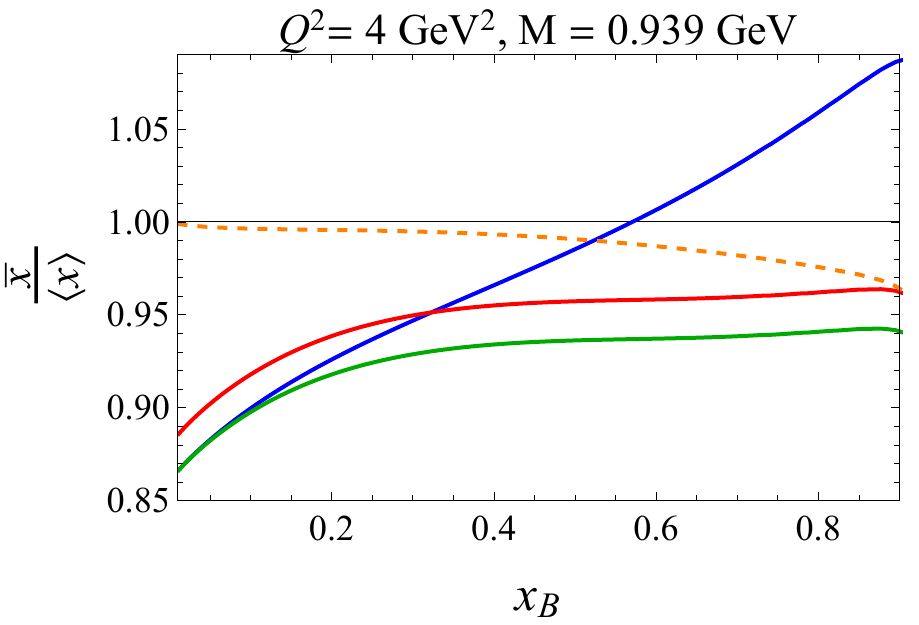}	
  \includegraphics[height=4.4cm]{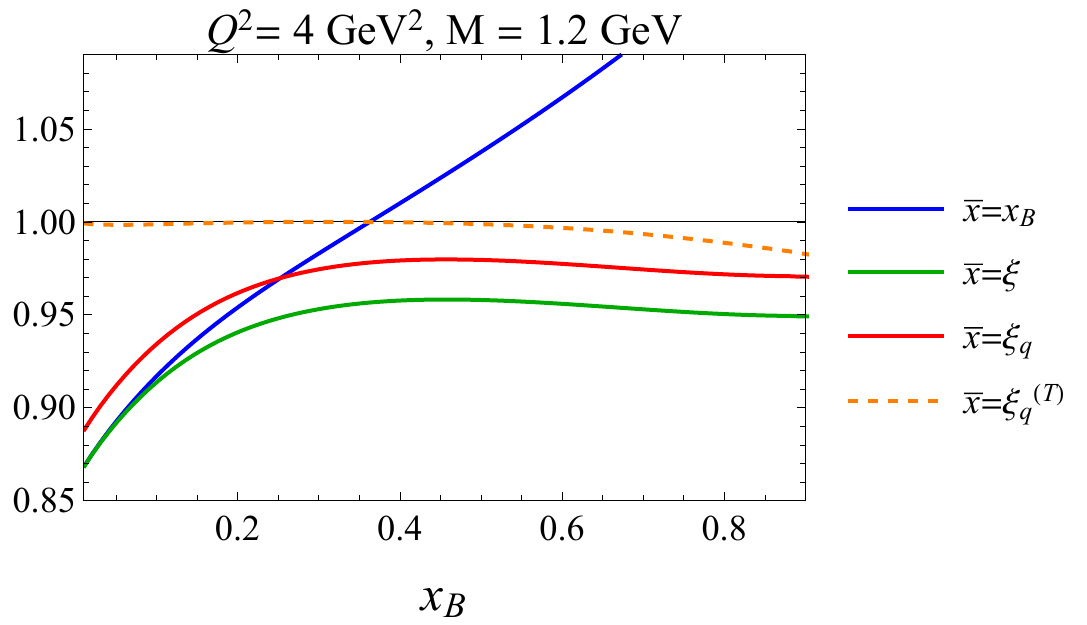}	
  \caption{Ratio of approximated $\bar x$ to the full $\langle x \rangle$ for two choices of target mass.}
  \label{Fig:avg_x}
\end{figure}

Finally, in Fig.~\ref{Fig:avg_x}, we compare the collinear $x=\bar x$ to the full $\langle x \rangle$. This reinforces the conclusion that $x=\xi_q$ is an adequate representation of the parton's longitudinal kinematics, but $x=x_B$ is not.

In conclusion, in the context of a spectator DIS model, we have verified that the range of validity of collinear factorization can be extended to subasymptotic values of $Q^2$ largely by using the mass-corrected scaling variable $\xi_q$ instead of $x_B$. We have also explicitly illustrated an inherent limitation of collinear factorization, that breaks down at large $x_B$ due to the neglect of momentum conservation in the transverse (and light-cone minus) direction.


\acknowledgments

This work was supported by the DOE contract No. DE-AC05-06OR23177, under which Jefferson Science
Associates, LLC operates Jefferson Lab, and  No. DE-SC0008791. J.V.G. was also partially supported
by the Jefferson Science Associates 2018-2019 Graduate Fellowship Program.

\bibliographystyle{JHEP}
\bibliography{DIS_2019} 

\providecommand{\href}[2]{#2}\begingroup\raggedright\begin{thebibliography}{1}

\bibitem{Accardi:2009md}
A.~Accardi, T.~Hobbs and W.~Melnitchouk, \emph{{Hadron mass corrections in
  semi-inclusive deep inelastic scattering}},
  \href{https://doi.org/10.1088/1126-6708/2009/11/084}{\emph{JHEP} {\bfseries
  11} (2009) 084} [\href{https://arxiv.org/abs/0907.2395}{{\ttfamily
  0907.2395}}].

\bibitem{Guerrero:2015wha}
J.~V. Guerrero, J.~J. Ethier, A.~Accardi, S.~W. Casper and W.~Melnitchouk,
  \emph{{Hadron mass corrections in semi-inclusive deep-inelastic scattering}},
  \href{https://doi.org/10.1007/JHEP09(2015)169}{\emph{JHEP} {\bfseries 09}
  (2015) 169} [\href{https://arxiv.org/abs/1505.02739}{{\ttfamily
  1505.02739}}].

\bibitem{Guerrero:2017yvf}
J.~V. Guerrero and A.~Accardi, \emph{{Gauge invariance and kaon production in
  deep inelastic scattering at low scales}},
  \href{https://doi.org/10.1103/PhysRevD.97.114012}{\emph{Phys. Rev.}
  {\bfseries D97} (2018) 114012}
  [\href{https://arxiv.org/abs/1711.04346}{{\ttfamily 1711.04346}}].

\bibitem{Bacchetta:2008af}
A.~Bacchetta, F.~Conti and M.~Radici, \emph{{Transverse-momentum distributions
  in a diquark spectator model}},
  \href{https://doi.org/10.1103/PhysRevD.78.074010}{\emph{Phys. Rev.}
  {\bfseries D78} (2008) 074010}
  [\href{https://arxiv.org/abs/0807.0323}{{\ttfamily 0807.0323}}].

\bibitem{Moffat:2017sha}
E.~Moffat, W.~Melnitchouk, T.~C. Rogers and N.~Sato, \emph{{What are the
  low-$Q$ and large-$x$ boundaries of collinear QCD factorization theorems?}},
  \href{https://doi.org/10.1103/PhysRevD.95.096008}{\emph{Phys. Rev.}
  {\bfseries D95} (2017) 096008}
  [\href{https://arxiv.org/abs/1702.03955}{{\ttfamily 1702.03955}}].

\bibitem{Aivazis:1993kh}
M.~A.~G. Aivazis, F.~I. Olness and W.-K. Tung, \emph{{Leptoproduction of heavy
  quarks. 1. General formalism and kinematics of charged current and neutral
  current production processes}},
  \href{https://doi.org/10.1103/PhysRevD.50.3085}{\emph{Phys. Rev.} {\bfseries
  D50} (1994) 3085} [\href{https://arxiv.org/abs/hep-ph/9312318}{{\ttfamily
  hep-ph/9312318}}].

\bibitem{Accardi:2008ne}
A.~Accardi and J.-W. Qiu, \emph{{Collinear factorization for deep inelastic
  scattering structure functions at large Bjorken x(B)}},
  \href{https://doi.org/10.1088/1126-6708/2008/07/090}{\emph{JHEP} {\bfseries
  07} (2008) 090} [\href{https://arxiv.org/abs/0805.1496}{{\ttfamily
  0805.1496}}].

\bibitem{Guerrero:2019xxx}
J.~V. Guerrero and A.~Accardi, \emph{{Collinear Factorization with hadron
  masses in a spectator model}}, {\emph{in preparation} }.

\bibitem{Qiu:1988dn}
J.-W. Qiu, \emph{{Twist Four Contributions to the Parton Structure Functions}},
  \href{https://doi.org/10.1103/PhysRevD.42.30}{\emph{Phys. Rev.} {\bfseries
  D42} (1990) 30}.

\end{thebibliography}\endgroup

\end{document}